\documentclass{aa}  %[referee]

\usepackage{graphicx}
\usepackage{lipsum}
\usepackage{txfonts}
\usepackage{color}
\usepackage{xcolor}
\usepackage{amsmath,amsfonts,amssymb}% Better maths support & more symbols
\usepackage{float}
\usepackage{hyperref}
\usepackage[version=3]{mhchem}
\usepackage{lineno}
\providecommand{\Msun}{M_{\odot}}
\providecommand{\Mjup}{M_{\textrm{Jup}}}
\providecommand{\masyr}{\ensuremath{\rm \,mas\,yr^{-1}}\xspace}

\begin{document} 
\title{Combining Hipparcos and Gaia data for the study of binaries: the BINARYS tool}

\author{A. Leclerc \inst{1}, C. Babusiaux \inst{1,3}, F. Arenou \inst{3}, F. van Leeuwen \inst{2}, M. Bonnefoy \inst{1}, X. Delfosse \inst{1}, T. Forveille \inst{1},  J.-B. Le Bouquin \inst{1}, L. Rodet \inst{4}}

   \institute{Univ. Grenoble Alpes, CNRS, IPAG, 38000 Grenoble, France \\
              \email{aurelia.leclerc$@$univ-grenoble-alpes.fr}
              \and
              Institute of Astronomy, University of Cambridge, Madingley Road, Cambridge CB3 0HA, UK
         \and
            GEPI, Observatoire de Paris, Université PSL, CNRS, 5 Place Jules Janssen, 92190 Meudon, France
         \and
         Cornell Center for Astrophysics and Planetary Science, Department of Astronomy, Cornell University, Ithaca, NY 14853, USA
        }

% \date{Received ° month 20.. ; Accepted ° month 20..}

% \abstract{}{}{}{}{} 
% 5 {} token are mandatory
 
  \abstract
   {Orbital motion in binary and planetary systems is the main source of precise stellar and planetary mass measurements, and joint analysis of data from multiple observational methods can both lift degeneracies and improve precision.}  
   {We set out to measure the masses of individual stars in binary systems using all the information brought by the Hipparcos and Gaia absolute astrometric missions.}
   {We present BINARYS, a tool which uses the Hipparcos and Gaia absolute astrometric data and combines it with relative astrometry and/or radial velocity measurements to determine the orbit of a binary system. It rigorously combines the Hipparcos and Gaia data (here EDR3), and it can use the Hipparcos Transit Data as needed for binaries where Hipparcos detect significant flux from the secondary component. It also support the case where Gaia resolved the system, giving an astrometric solution for both components.}
   {We determine model-independent individual masses for the first time for three systems: the two mature binaries Gl~494 ($M_1=0.584 \pm 0.003 \Msun$ and $M_2=87 \pm 1 \Mjup$) and HIP~88745 ($M_1=0.96 \pm 0.02 \Msun$ and $M_2= 0.60^{+ 0.02 } _{- 0.01 } \Msun$), and the younger AB Dor member GJ~2060 ($M_1=0.60 ^{+ 0.06} _{- 0.05} \Msun$ and $M_2=0.45 ^{+ 0.06} _{- 0.05}\Msun$). The latter provides a rare test of evolutionary model predictions at young ages in the low stellar-mass range and sets a lower age limit of 100~Myr for the moving group.}
    {}

   \authorrunning{A. Leclerc et al.}
   \maketitle
%
%-------------------------------------------------------------------
\section{Introduction}

  The study of binaries is a constantly expanding field of research and the combination of multiple observational methods for their characterization is frequently used because it allows to determine directly the masses of each component. Absolute astrometry has a very long history as a method to identify and study invisible stellar, and more recently planetary, companions to stars. It has however up to now been applicable to fairly small samples. The ongoing Gaia mission \citep{GaiaMission} represents, in this respect as in many others, a game changer, and is expected to astrometrically detect thousands of planets and stellar companions. Its epoch data, however, will only become available with the 4$^{th}$ Gaia data release \footnote{https://www.cosmos.esa.int/web/gaia/release}. Until then, the data of Gaia predecessor Hipparcos \citep{HIP1} provide a very valuable test-bed, since both missions similarly obtain astrometry along a single direction. Moreover, combining Hipparcos and Gaia data extends the time base of the measurements to over three decades, which will always be essential for longer period binaries.

Several methods have already been used to mine the multiplicity information brought by the combination of Hipparcos and Gaia. The earliest has been comparison of the "short-term" proper motions returned by both missions against the "long-term" proper motion derived from the difference between the Hipparcos and Gaia positions \citep[see][for a review of this method]{Kervella2019}.
This produced both astrometric acceleration catalogues \citep{Brandt2018,Kervella2019,Brandt21,Kervella22} and statistical rough companion mass determinations \citep{Kervella2019}. More recently, \cite{orvara2021} and \cite{htof2021} developed tools that adjust Gaia \citep{orvara2021} information together with respectively radial velocities and relative astrometric data \citep{orvara2021}, and Hipparcos Intermediate Astrometric Data \citep[IAD,][Vol. 1, Sect.~2.8]{HIP1}. Much earlier, \cite{Soderhjelm1999} combined the raw Hipparcos data, called Transit Data \citep[TD,][Vol. 1, Sect.~2.9]{HIP1}, from the original Hipparcos reduction \citep{HIP1} with ground-based observations to adjust orbital elements, but no attempt has to date been made to combine those with Gaia information.

BINARYS (orBIt determiNAtion with Absolute and Relative astrometRY and Spectroscopy) is our new tool and adjusts together: the residual abscissae from Hipparcos data (IAD or TD), the astrometric parameters available from Gaia, and complementary observations from relative astrometry and radial velocity. BINARYS uses a gradient descent method implementing automatic differentiation thanks to the R package TMB (\citealp{TMB}), and it 
rigorously uses the information from Hipparcos and Gaia with minimal assumptions or simplifications. 

In the following we first present in Sect.\ref{section:Data} the data classes used by BINARYS, and then in Sect.\ref{section:method} the tool itself as well as its limits and validation. In Sect.\ref{section:results}, we illustrate the capabilities of BINARYS with 3 systems: Gl~494 where we combine relative astrometry with the Hipparcos IAD and Gaia data, GJ~2060 which we analyse with relative astrometry and Hipparcos TD, and HIP~88745 where we combine Hipparcos TD with Gaia resolved observations.

%%%%%%%%%%%%%%%%%%%%%%%%%%%%%%%%%%%%%%%%%%%%%%%%%%%%%%%%%%%

\section{The data}
\label{section:Data}

\subsection{Hipparcos}

The Hipparcos mission \citep{HIP1} operated from 1989 to 1993. The satellite scanned the sky continuously along great circles, and projected the image of pre-selected stars through an alternatively transparent and opaque grid with grid step $s=1207.4$ mas which modulated their light. The one-dimensional (1D) position of the object along the scanning great circle is thus encoded into the observed phase of the corresponding quasi-periodic signal. Each star was observed during multiple satellite transits and the observations are published as residual abscissae (noted $\Delta \nu$), which are the difference between the observed position of the star and the predicted position along the scanning circle for the published best model of the star. Two data reductions are available: the original reduction (\citealp{HIP1}) and the new reduction (\citealp{HIP2}). The tool can handle both reductions, but in the following sections we will only use data from the new reduction.

The residual abscissae are published in the Intermediate Astrometric Data (IAD)\footnote{\url{https://www.cosmos.esa.int/documents/532822/6470227/ResRec_JavaTool_2014.zip}. 
We note that \cite{htof2021} found an issue on the IAD which is that the astrometric solutions obtained from them are not exactly the same as the published ones when the number of observations NOB is lower than the number of residual records NRES. We confirm this issue but it does not seem to be due to data corruption as we do not find repeated sequences for the along-scan errors described in \cite{htof2021}.
 
We checked that no source studied here is affected.} and can be used when the observed object is a point-source for Hipparcos. When the object instead is a resolved binary or multiple system, the observed phase no longer measures its photocentre, but instead something specific to the Hipparcos scanning grid method and which has been dubbed the Hippacentre \citep{Martin1997}. 
Using only the IAD, it is possible to use the Hippacentre to constrain the mass and the intensity ratio of the components, as shown by \cite{Martin1997}. However, the transit Data (TD) contains the full signal modulation parameters and therefore provide more constraints on the Hipparcos observations of a resolved system \citep{Quist1999}. \cite{Soderhjelm1999} pioneered the usage of the TD to derive masses of visual binaries. While in the original Hipparcos solution only a third of the sources have their TD provided, in the reduction of \cite{HIP2} the TDs are available for all Hipparcos stars.

Those TD are extracted by the Java Tool\footnote{\url{https://www.cosmos.esa.int/web/Hipparcos/interactive-data-access}. An ASCII version of the TD are in preparation as well as an update of the tool to retrieve those data.
The TD files provided in the DVD suffer for some stars from a factor 10 issue on the $\beta_5$ and associated error and should therefore not be used.
} from the Hipparcos calibrated raw data. The new reduction did not re-process the photometric signal which should then be retrieved in the original reduction Epoch Photometric Annex and Extension (accessible via ESASky legacy TAP query \footnote{\url{https://www.cosmos.esa.int/web/esdc/esasky-catalogues}: \texttt{hipparcos1.hip\_ep} and \texttt{hipparcos1.hip\_ep\_e}.}). 
Some transits do not have photometric information available in the original reduction and we ignore those transits in our code.

For the original reduction, BINARYS can handle either the IAD, which list the residual abscissae relative to the published 5-parameters astrometric model, or the TD for which we apply the method described in \cite{Quist1999}. For the new reduction, the residual abscissae are given relative to the model used for each star, which need not be the 5-parameters model. BINARYS only uses the IAD when the solution for the star is a 5-parameters solution, and reverts to using the TDs when it was analysed with a different model.\footnote{Note that a small difference between the IAD and TD abscissae residuals are present due to the different handling of $\beta_5$ (see Sect.\ref{subsection:hippa}): $\Delta\nu^5_{IAD} = \Delta\nu^5_{TD} -  11.5356 \, \beta_5$.}

\subsection{Gaia EDR3}
The Gaia mission \citep{GaiaMission} started observing in 2014 and is ongoing. The observation principle of Gaia is similar to Hipparcos, except that the satellite records small images rather than a periodically modulated signal. The raw data are not available yet and the different data releases to date (DR1: \citealp{DR1}, DR2: \citealp{DR2} and EDR3: \citealp{eDR3}) only provide the 2 or 5 astrometric parameters that best match the observations, without taking into account a possible multiplicity. Multiplicity will be taken into account for the first time in the forthcoming DR3 release.

\par The scanning law of the satellite, which contains the pointing direction and the scanning angle as a function of time, is also published\footnote{\url{https://gea.esac.esa.int/archive/documentation/GEDR3/Gaia\_archive/chap\_datamodel/sec\_dm\_auxiliary\_tables/ssec\_dm\_commanded\_scan\_law.html}} and provides the conditions under which a given star was observed. 

Before using the Gaia data, we also examine ancillary information such as the \texttt{ruwe} (Renormalised Unit Weight Error) and the multi peak flag (\texttt{ipd\_frac\_multi\_peak}) present in EDR3 \citep{Lindegren21a}. The \texttt{ruwe} evaluates the quality of the 5-parameter solution, and a value above 1.4 indicates that the published solution may %does 
not describe the object well \citep{LL-124}. The multi-peak flag indicates the percentage of the windows used for the astrometric processing of the source which contain a double peak, and a high value is evidence of flux contamination. For us to use the Gaia data, the signal must originate from the source alone, since accounting for flux contamination would require a model of the line spread function fitting which is not published at this point. When Gaia does not fully resolve the system and the secondary contributes non negligible light, we cannot use the Gaia data, which corresponds to separations between 9~mas and 0.27\arcsec depending on the magnitude difference \citep{LL-136}. For smaller separations the photocentre can be used.
If Gaia fully resolves the system and gives a separate solution for each component, then BINARYS can use those solutions, even when they are just 2-parameter solutions which ignore parallax and proper motion. Analysis of partially resolved systems with non negligible light from companions will have to wait for more detailed Gaia data, which will become available in the DR4 release.

\par To combine Gaia and Hipparcos data, we have to bring them into the same reference frame. We somewhat arbitrarily chose to convert the Gaia positions and proper motions to the Hipparcos proper motion reference frame (\citealp{rotPrinciple}, \citealp{Brandt2018}, \citealp{Kervella2019}). As a consequence, the astrometric parameters which BINARYS adjusts to the data are in the Hipparcos reference frame at epoch: $\mathrm{HIP}_\mathrm{epoch}= J1991.25$. 
 The rotation to be used for Gaia EDR3 is $\Omega_{GH} = \left[\begin{smallmatrix} \omega_X \\ \omega_Y \\ \omega_Z \end{smallmatrix}\right] = \left[\begin{smallmatrix} -0.120 \\ 0.173 \\ 0.090 \end{smallmatrix}\right] \masyr $ \citep{rotValeDR3}, and the transformed Gaia astrometric parameters are given by:
\begin{equation}
\begin{aligned}
 \left[\begin{smallmatrix} \alpha_{new}^* \\ \delta_{new} \end{smallmatrix}\right] 
& = \left[\begin{smallmatrix} \alpha^* \\ \delta \end{smallmatrix}\right] +
 A\cdot\Omega_{GH}\;  \Delta_{GH} \;  \\
  \left[\begin{smallmatrix} \mu_{\alpha_{new}^*} \\ \mu_{\delta_{new}} \end{smallmatrix}\right] 
& = \left[\begin{smallmatrix} \mu_{\alpha^*} \\ \mu_\delta \end{smallmatrix}\right] +
 A\cdot\Omega_{GH} 
\end{aligned}
\end{equation}
with $\alpha^*=\alpha\,\cos{\delta}$, $\mu_{\alpha^*}=\mu_\alpha\,\cos{\delta}$ and the polar to Cartesian coordinates transformation matrix:
\begin{equation}
A= \left[ \begin{smallmatrix} \cos\alpha\sin\delta & \sin\alpha\sin\delta & -\cos\delta \\ -\sin\alpha & \cos\alpha & 0 & \end{smallmatrix} \right] \\
\end{equation}
and with $\Delta_{GH}$ the difference between the Gaia ($\mathrm{EDR3}_\mathrm{epoch}= J2016$) and $\mathrm{HIP}_\mathrm{epoch}$ epochs. 
Similarly, we correct for the parallax zero point difference: $\varpi_{new}= \varpi - \varpi^{shift}$, with $\varpi^{shift}_{EDR3}=-0.068$ mas \citep{rotValeDR3} after applying the Gaia EDR3 parallax correction proposed by \cite{Lindegren21b}.
The uncertainties on the 5 astrometric parameters are inflated according to the parallax error under-estimation factor derived by \cite{ElBadry21}.

\subsection{Relative astrometry}
Relative astrometry data can originate from either direct imaging or interferometric observations and it consists of relative positions of the components at one or several epochs. The inputs for BINARYS are the date of the observation and the relative position of the two components in $\alpha$ and $\delta$ direction ($\xi$ and $\eta$) with their associated uncertainties ($\sigma_\xi$ and $\sigma_\eta$).

When the relative positions are published as a separation and position angle ($\rho$ and $\theta$) and unless the publication includes a full covariance matrix, we adopt as covariance matrix $\Sigma_{\xi\eta}$:
\begin{equation}
\begin{aligned}
& \Sigma_{\xi\eta} = J \cdot \Sigma_{\rho\theta} \cdot J^{T} \; \; \;  \Sigma_{\rho\theta} = \left[ \begin{smallmatrix} \sigma_\rho & \mathrm{cor}(\rho,\theta) \\ \mathrm{cor}(\rho,\theta)  &\sigma_\theta & \end{smallmatrix} \right] \; \; \; J = \left[ \begin{smallmatrix} \sin\theta & \rho\cos\theta \\ \cos\theta & -\rho\sin\theta  \end{smallmatrix} \right]
\end{aligned}
\end{equation}
with J the Jacobian of the polar to Cartesian transformation and $cor(\rho,\theta)$ assumed null. 

\subsection{Radial velocity}

The radial velocity inputs contain the date of the observation, the radial velocity (RV), its uncertainty,
and optionally a code for the instrument which was used.
The latter is needed when the radial velocity inputs were obtained with multiple instruments, and allows to adjust offsets to account for RV zero point differences. A jitter can also be added, either to increase the instrument noise or to take into account an un-modelled stellar variability. The radial velocities can be adjusted for either the primary or the secondary stars, allowing to handle both single-lined and double-lined binaries.

%%%%%%%%%%%%%%%%%%%%%%%%%%%%%%%%%%%%%%%%%%%%%%%%%%%%%%%%%%%%
\section{Method: combination of absolute astrometry with relative astrometry and radial velocity}
\label{section:method}

In the following we describe how BINARYS estimates the orbital (OP) and astrometric (AP) parameters of a binary system.

The adjusted OP are expressed using the Campbell elements $\theta_{OP}=\{P,T_p^{rel},a_1,e,i,\omega_1,\Omega,X\}$, where $X$ can be either $\{a_{21}\}$ or $\{M_1\}$. $P$ is the period in years, $T_p^{rel}$ defines the epoch of one periastron, counted from J2000.0 in units of the orbital period ($T_P(J2000)= T_p^{rel} \, P$). $a_1$ and $a_{21}$ are the semi major axis of respectively the orbit of the primary and the relative orbit, in au. $e$ is the eccentricity, and $i$ is the inclination of the orbital plane to the tangent plane of the sky, oriented with the convention that $0 \leq i \leq 90^{\circ}$ for a direct (defined by an increasing positional angle that is counted positive from north towards east direction) apparent motion and $90 \leq i \leq 180^{\circ}$ for a retrograde apparent motion. $\omega_1$ is the argument of periastron of the primary, counted from the ascending node and in the direction of the motion. The argument of periastron of the secondary is linked with that of the primary by $\omega_2=\omega_1 + \pi$. $\Omega$ is the position angle of the ascending node, with the conventions used in Hipparcos and Gaia: it is the position angle, counted counterclockwise from the $\delta$ direction, of the intersection of the orbital and tangent planes. When radial velocities are available and resolve the ambiguity between the two nodes, $\Omega$ corresponds to the node where the primary star recedes from the observer; otherwise, we arbitrarily impose $0 \leq \Omega \leq 180^{\circ}$. Finally, $M_1$ is the mass of the primary in units of solar masses.
When combining absolute astrometry with relative astrometry, fitting $a_{21}$ is a natural choice while combining absolute astrometry with radial velocities is easier using $M_1$, parameter which has the advantage of having spectroscopic and/or photometric estimates. Both parameters lead in practice to the mass ratio information $q=M_2/M_1$ through the equations:
\begin{equation}
M_1 (1+q) = \frac{a_{21}^3}{P^2}    \\ \text{and } \\
a_{21} = a_1 (1+\frac{1}{q})
\end{equation}
To handle the photocentre motion, BINARYS also adjusts the fractional luminosity $\beta=\frac{L_2}{L_1+L_2}$.
The adjusted astrometric parameters are the usual $\theta_{AP}=\{\alpha,\delta,\varpi,\mu_{\alpha^*},\mu_\delta\}$. 

\subsection{Adjustment of relative astrometry data}
\label{relastrom}

For a given observation time $t$, the positions of the primary (1) and secondary (2) stars, relative to the barycentre and along the $\alpha$ and $\delta$ direction ($\xi$ and $\eta$), are computed as:
\begin{equation}
\label{relatpos}
\begin{aligned}
&{\xi}_1 =  D \left( \cos(\upsilon + \omega_1)\sin{\Omega} + \sin(\upsilon + \omega_1)\cos{\Omega}\cos{i} \right) \\
&{\eta}_1= D \left( \cos(\upsilon + \omega_1)\cos{\Omega} - \sin(\upsilon + \omega_1)\sin{\Omega}\cos{i} \right)\\
&{\xi}_2 = \frac{D}{q} \left( \cos(\upsilon + \omega_2)\sin{\Omega} + \sin(\upsilon + \omega_2)\cos{\Omega}\cos{i} \right) \\
&{\eta}_2=\frac{D}{q} \left( \cos(\upsilon + \omega_2)\cos{\Omega} - \sin(\upsilon + \omega_2)\sin{\Omega}\cos{i} \right)
\end{aligned}
\end{equation}
with the polar coordinates of the primary on its orbit $D= a_1 \frac{1-e^2}{1+e \cos{\upsilon}}$ and  
%with $D = a_1 \times \left[1 - e \times \cos(E)\right]$
%, $q$ the mass ratio (=$M_2/M_1$), 
$\upsilon$ the true anomaly. $\upsilon$ is related to the eccentric anomaly $E$ through
\begin{equation}
    \tan{\frac{\upsilon}{2}}=\sqrt{\frac{1+e}{1-e}}\tan{\frac{E}{2}}
\end{equation}
and $E$ is obtained by numerically solving Kepler's equation 
\begin{equation}
2 \pi (t-T_p)/P = E - e \sin{E}     
\end{equation}
over 10 iterations \citep{Heintz78}. 
The relative positions between the two stars, in au units, are obtained by differencing as $\Delta\xi =  {\xi}_2 - {\xi}_1$ and $\Delta\eta=  {\eta}_2 - {\eta}_1$, and converted to angular separations by multiplying with the parallax (which is one of the astrometric parameters). We finally compute the residuals between the computed and observed relative positions.

\subsection{Adjustment of radial velocity data}
To predict the radial velocity, we first calculate from the orbital parameters the semi amplitude $K$ (km/s) of the radial velocity signal as, for the primary and secondary:
\begin{equation}
\begin{aligned}
&K_{1} = C \frac{a_{1} \sin{i}}{P \sqrt{1 - e^2}}\\
&K_{2} = \frac{K_1}{q}
\end{aligned}
\end{equation}
with C = 29.78525 km/s ($= \frac{2\pi \, AU}{365.25\times24\times3600}$) and $AU$ the astronomical unit in kilometres. The predicted radial velocity for a given epoch is then, for the primary and the secondary:
\begin{equation}
\begin{aligned}
&RV_1 = RV_0 + K_1 \left[cos(\upsilon + \omega_1) +e\cos \omega_1\right]\\
&RV_2 = RV_0 + K_2\left[cos(\upsilon + \omega_2)+e\cos \omega_2 \right]
\end{aligned}
\end{equation}
with $RV_0$ the radial velocity of the barycentre to be adjusted. The predicted radial velocity is then compared to the observed radial velocity data.% thanks to a likelihood maximization.

\subsection{Adjustment of Hipparcos data}
\label{subsection:hippa}

To predict the residual abscissae $\Delta \nu$ and compare them to the observed ones, we first have, at each observing epoch, to project the separation of the two stars on the Hipparcos scanning grid, for the orientation of the grid at that epoch. The separation of the components along $\alpha$ and $\delta$ are the $\Delta\xi$ and $\Delta\eta$ calculated in Sect. \ref{relastrom}. To project them on the grid, we resort to the partial derivatives of the abscissa against the 5 astrometric parameters $\frac{\partial \nu }{\partial a_j}$ with $a_j=\{\alpha, \delta, \varpi, \mu_{\alpha^*}, \mu_\delta\}$, which are:
\begin{equation}
\begin{aligned}
&\frac{\partial \nu }{\partial \alpha} = \cos{\psi}  \; \; ; \; \; \frac{\partial \nu }{\partial \delta} = \sin{\psi} \; \; ; \; \; \frac{\partial \nu }{\partial \varpi} = \varpi_{factor}\\
&\frac{\partial \nu }{\partial \mu_{\alpha^*}} =  \cos{\psi}\, \Delta T  \; \; ; \; \;\frac{\partial \nu }{\partial \mu_\delta} = \sin{\psi}\,  \Delta T
\end{aligned}
\end{equation}
with $\psi$ the position angle of the scanning direction, $\varpi_{factor}$ the parallax factor (\citealp{HIP1}, Eq. 1.2.26) and $\Delta T$ the observation epoch relative to $HIP_{epoch}$. Those are available in the IAD and the TD of the new Hipparcos reduction, while the original reduction directly provides the 5 partial derivatives.\\

For each transit and for a given set of orbital parameters $\theta_{OP}$, the projected separation $\rho_{p}$ on the scanning direction is:
\begin{equation}
\rho_p = ({\xi}_{2}-{\xi}_{1})\, \cos{\psi} + ({\eta}_{2}-{\eta}_{1})\, \sin{\psi}
\label{rhop}
\end{equation}
%^{\theta_{OP}}
We then calculate $\zeta$, the projected separation in units of the Hipparcos grid step:
\begin{equation}
\zeta = 2\pi\rho_{p}/s.
\end{equation}

Finally, we calculate $\Delta\nu_B$, the position shift along the scanning direction due to the binary motion:
\begin{equation}
\Delta\nu_B = \frac{\phi\,s}{2\pi} - B\, \rho_{p}
\label{nuB}
\end{equation}
with B the fractional mass, or in other words the mass of the secondary divided by the total mass of the system $B=\frac{M_2}{M_1+M_2}$, and the phase $\phi = \mathrm{atan2}(\beta \sin{\zeta}, 1-\beta+\beta\cos{\zeta})$ where $\beta$ is the fractional luminosity $\beta = (1+10^{0.4\Delta m_{HIP}})^{-1}$. For a detailed explanation of those steps, see \cite{Martin1997} who used the function $\mathrm{Angle}(x,y)$, equivalent to $\mathrm{atan2}(y,x)$. 
If the flux of the secondary is negligible, $\beta$ becomes zero and the $\Delta\nu_B$ shifts are purely from the reflex orbit of the primary star. If the secondary star contributes light but the separation stays small compared to the grid step, the orbit is that of the photocentre.\\

We then calculate $\Delta\nu$ by applying the partial derivatives to the difference between the 5 adjusted astrometric parameters ($\theta_{AP}$) and the 5 published reference astrometric parameters ($\widehat{\theta_{AP}}$):
\begin{equation}
\Delta\nu = \Delta\nu_B+ \sum_{j=1}^5 (a_j^{\theta_{AP}} - a_j^{\widehat{\theta_{AP}}}) \frac{\partial\nu}{\partial a_j}
\end{equation}

These new computed residuals are then compared to the observed ones for each transit.\\

\par When neither the secondary flux nor the separation is negligible, we include in the adjustment 3 additional observational quantities available in the Hipparcos transit data \citep{TD2book}, $\beta_4$, $\beta_5$, and $H_{AC}H_{DC}=Hp_{ac}-Hp_{dc}$. $\beta_4$ and $\beta_5$ describe the amplitude and the phase of the second harmonic of the grid-modulated signal and are closely related to an interferometric visibility measured at the corresponding angular frequency. $Hp_{dc}$ and $Hp_{ac}$ are magnitudes (in the $Hp$ Hipparcos spectral bandpass) evaluated from, respectively, the unmodulated DC and the modulated AC components of the Hipparcos transit signal. These three observable quantities can be computed from the adjustable parameters as:
\begin{equation}
\begin{aligned}
&\beta_{4} = (1+(r+r^2)\,(2 \cos{\zeta}+\cos{2\zeta})+r^3)/nf \\
&\beta_{5} = (r-r^2)\,(2 \sin{\zeta}-\sin{2\zeta})/nf\\
&H_{AC}{H_{DC}} = -2.5\log_{10}\left(\sqrt{1+2r\cos{\zeta}+r^2}/(1+r)\right)
\end{aligned}
\end{equation}
with $r$ the ratio of the secondary and primary luminosities, which can be written as $r=\beta/(1-\beta)$, and $nf=\left(1+2r\cos{\zeta}+r^2 \right)^{3/2}$.

The uncertainties on $\Delta\nu$ are modified according to the amplitude of the first harmonic of the modulated signal, which for a binary is decreased from its point-source value by a factor $f_{\sigma_\nu} = (1+r)/\sqrt{1+2r\cos{\zeta}+r^2}$. When comparing the computed and observed $\Delta\nu$, we consequently have to increase the measurements errors $\sigma_{\Delta\nu}$ by the factor $f_{\sigma_\nu}$. \\

When one of the components is variable, we adjust one value of the fractional luminosity $\beta$ for each epoch rather than one common value. $\beta$ is then calculated as $\beta = \frac{r}{1+r}$, with $r=r_1$ when the primary is variable, and $r=r_2$ when the secondary is variable:
\begin{equation}
\begin{aligned}
& r_1 =  \left(\frac{1}{\beta_0} \, 10^{-0.4 \, (Hp_{dc}-\overline{Hp_{dc}})}-1\right)^{-1} \\
& r_2 = \frac{1}{1-\beta_0}\, 10^{-0.4\,(Hp_{dc}-\overline{Hp_{dc}})} - 1
\end{aligned}
\end{equation}
with $\beta_0$ and $\overline{Hp_{dc}}$ respectively the mean fractional luminosity and total magnitude \citep{HIP2}.

\subsection{Adjustment of Gaia data}
\label{subsection:gaia} 

The Gaia observations also constrain the astrometric and orbital information but, despite the impressive size of the Gaia DR3 non-single star catalogue \citep{Arenou22}, for the majority of the systems, we only have access to the set of 5 astrometric parameters as those systems were not yet treated as binaries by Gaia DR3. For those systems, the Gaia orbital information is therefore encoded in the bias of the 5 published astrometric parameters away from their true barycentric values.
We therefore proceed by computing what Gaia would have observed for a given set of orbital and barycentric astrometric parameters and comparing to the 5 "effective" astrometric parameters published in the [E]DR3 catalogue. 

To do that, we start by propagating the astrometric parameters of the barycentre to the mean epoch of Gaia DR3. For nearby stars, that transformation must take the radial velocity into account (\citealp{HIP1}: Sect. 1.5). 
Perspective acceleration during the Hipparcos and Gaia observations, only needed for the closest stars, are not yet taken into account.
We then retrieve from the published Gaia scanning law the epochs when one of the two Gaia fields of view passed over the target of interest, as well as the scanning angle $\psi$ for each of those epochs. We assume in the following that all those epochs have contributed to the Gaia solution.

We proceed to compute, for each epoch, the orbital motion projected along the Gaia scanning direction. When Gaia does not resolve the system, that motion is that of the photocentre. If instead Gaia resolved the system and gives separate solutions for the two components, we compute individual offsets for the primary and the secondary stars.
The positions relative to the barycentre for the photocentre (0), the primary (1) and the secondary (2) projected along the Gaia scanning direction are respectively:

\begin{equation}
\label{gaiares}
\begin{aligned}
&\Delta\nu_B^{G, 0} = \left( (\xi_{2}-\xi_{1})\, \cos{\psi}   + (\eta_{2}-\eta_{1})\, \sin{\psi}  \right) \, (\beta-B) \\
&\Delta\nu_B^{G, 1} = \xi_{1}\, \cos{\psi}   + \eta_{1}\, \sin{\psi}  \\
&\Delta\nu_B^{G, 2} =  \xi_{2} \,\cos{\psi}   + \eta_{2}\, \sin{\psi} 
\end{aligned}
%\label{eqGaiaPos}
\end{equation}
The first equation in Eq. \ref{gaiares} for Gaia (also writable as $\Delta\nu_B^{G, 0} = (\beta - B) \; \rho_p$) is similar 
to Eq. \ref{nuB} for Hipparcos except that the extra complication in the Hipparcos formulation comes from the fact that Eq. \ref{nuB} corresponds to the Hippacentre, instead of the photocentre, due to the signal modulation \citep{Martin1997}.
From those abscissa residuals along the Gaia scanning direction for each observing epoch and the astrometric parameters propagated to the mean Gaia epoch $AP^G_0$, we estimate the 5 astrometric parameters that Gaia would have observed $AP^G$ and compare those with the published parameters:
\begin{equation}
AP^G = AP^G_0 + X
\end{equation}
with X the variation in astrometric parameters that reflects the residuals due to the binarity. It is obtained by solving the linear equation: $R = D \cdot X$, with $R$ the matrix of residuals $\Delta\nu_B$ and $D$ the matrix of partial derivatives \citep[Sect. 3.1]{vanLeeuwenEvans98}. 
The partial derivative along-scan $\varpi_{factor}$ is computed using the position of the system on the sky, the observation epoch and the corresponding scanning angle, as well as the orbit of the Earth (\citealp{HIP1}, Eq. 1.2.26). 
For targets where Gaia published only positions and no proper motion and parallax, we nonetheless compute all 5 astrometric parameters and then discard the parallax and the proper motion. 
In the Gaia processing, a Galactic prior information is added to provide more realistic uncertainties, which is not needed to mimic here.

\subsection{TMB: source code, options and limitations of the tool}

TMB (Template Model Builder,\citealp{TMB}) is an open source R package designed to quickly and robustly adjust non-linear models with large number of parameters. The R code calls functions from a user-provided C++ file that compute the likelihoods, which we make available \footnote{\url{https://gricad-gitlab.univ-grenoble-alpes.fr/ipag-public/gaia/binarys}}.
To briefly explore the uncertainties and degeneracies of the parameters, we post-process the TMB results with a short MCMC run using the companion R package \texttt{tmbstan} \citep{tmbstan}. In the present paper the orbits are well constrained and a single short MCMC chain of 3000 iterations of which we discard the first 1500 as warm-up iterations has been found to be enough to reach convergence.

To help TMB converge, we adopt starting values from the literature whenever available. For previously unstudied systems, we explore a large range of starting values and often initially fix some parameters to plausible values (e.g. starting with a circular orbit or fixing the primary mass).
When adjusting to Hipparcos transit data, good starting values for the astrometric parameters that already take into account a preliminary orbit greatly help. An option to ignore $f_{\sigma_\nu}$ for the first few iterations can also help to quickly obtain starting values for the astrometric parameters.
When adjusting for a radial velocity jitter, its value is best determined through a MCMC run and then set fixed when running TMB, because MCMC is less disturbed by jitter than gradient descent algorithms.
When the flux ratio of the two stars is available, whether from Gaia or from ancillary observations through similar filters, it can enter the adjustment as an observation with its uncertainty. 

The TMB adjustment works for well constrained orbit but needs enough data. 
The tool takes into account the system's perspective acceleration between the Hipparcos and Gaia epochs, but not along the Gaia mission as needed for very nearby and/or fast moving stars. Also, for stars identified by Hipparcos with a component solution, it can happen that the light that has contaminated the
data does not come from a companion of the system but from another star as it can happen in clusters. Both issues are matter for future developments.

\subsection{The evaluation of the solution}
\label{subsection:eval}

We evaluate each adjustment through its goodness of fit F2 \citep{F2}, which asymptotically follows a Gaussian distribution and which is defined as: 
\begin{equation}
F2 = \sqrt{\frac{9 \, k}{2}} \left[ \left(\frac{\chi_\mathrm{tot}^2}{k}\right)^{1/3} + \frac{2}{9 \, k} - 1\right]
\end{equation}
with $k$ the number of degrees of freedom (the number of observations minus the number of adjusted parameters) and $\chi_\mathrm{tot}^2$ the sum of the $\chi^2$ contributions of the individual observational methods. To be qualified as good, the adjustment must have F2 below 3.\\

To test the improvement of our solution on the Hipparcos data,
we compute, as in Hipparcos, an F2 using only the $\chi^2$ contribution associated with the residual abscissa (with their uncertainties increased by the $f_{\sigma_\nu}$ multiplicative factor described in the Sect.\ref{subsection:hippa}), and the number of parameters adjusted for the published solution. For the global F2 of our adjustments, $\beta_4$, $\beta_5$ and $H_{AC}H_{DC}$ also contribute to the Hipparcos $\chi^2$.

\section{Orbital study of benchmark systems}
\label{section:results}
We choose for illustration below three binary systems which use different data type combinations and which also provide interesting astrophysical results. The Gl~494 system tests the combination of Hipparcos IAD with Gaia astrometric parameters and relative astrometry, GJ~2060 tests the combination of Hipparcos TD with relative astrometry (including a new GRAVITY observation, Table~\ref{GRAVITYpoint}), and HIP~88745 tests the combination of Hipparcos TD with Gaia resolved observation. For these three stars, radial velocity data are also available but were not included in the adjustment: we only used them as an independent check of the results of the adjustments.

\subsection{Gl~494: Hipparcos Intermediate Astrometric Data, Gaia and direct imaging}
\label{sec:GL494}

\begin{figure*}[htp]
	\begin{minipage}[t]{0.48\linewidth}
		\centering
		\includegraphics[scale=0.30]{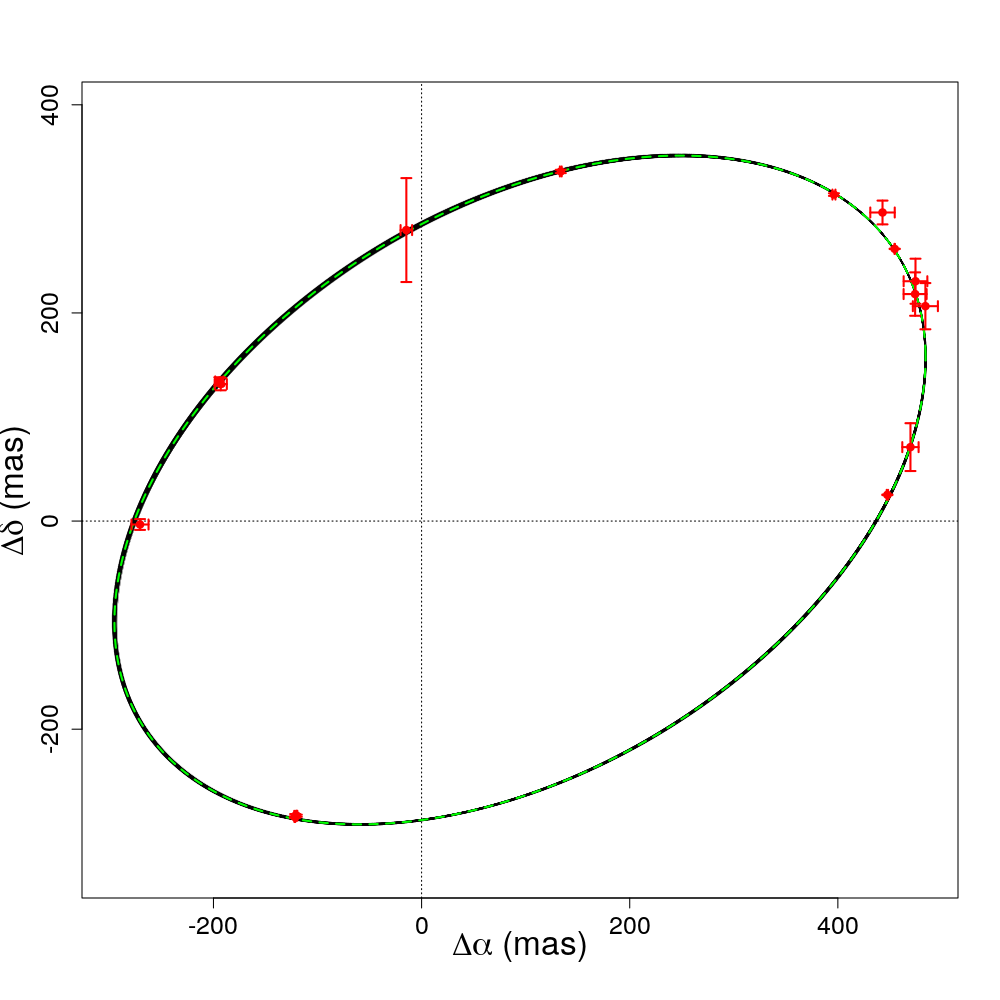}
		\caption{Gl~494 best fitting orbit from TMB (values in Table~\ref{OPTable}) as the green dotted line and sample orbits from the MCMC algorithm in black. The direct imaging observations \citep{Bowler2020} with their associated error bars are in red.}
		\label{GL494}
	\end{minipage}
	\hfill
	\begin{minipage}[t]{0.48\linewidth}
		\centering
		\includegraphics[scale=0.30]{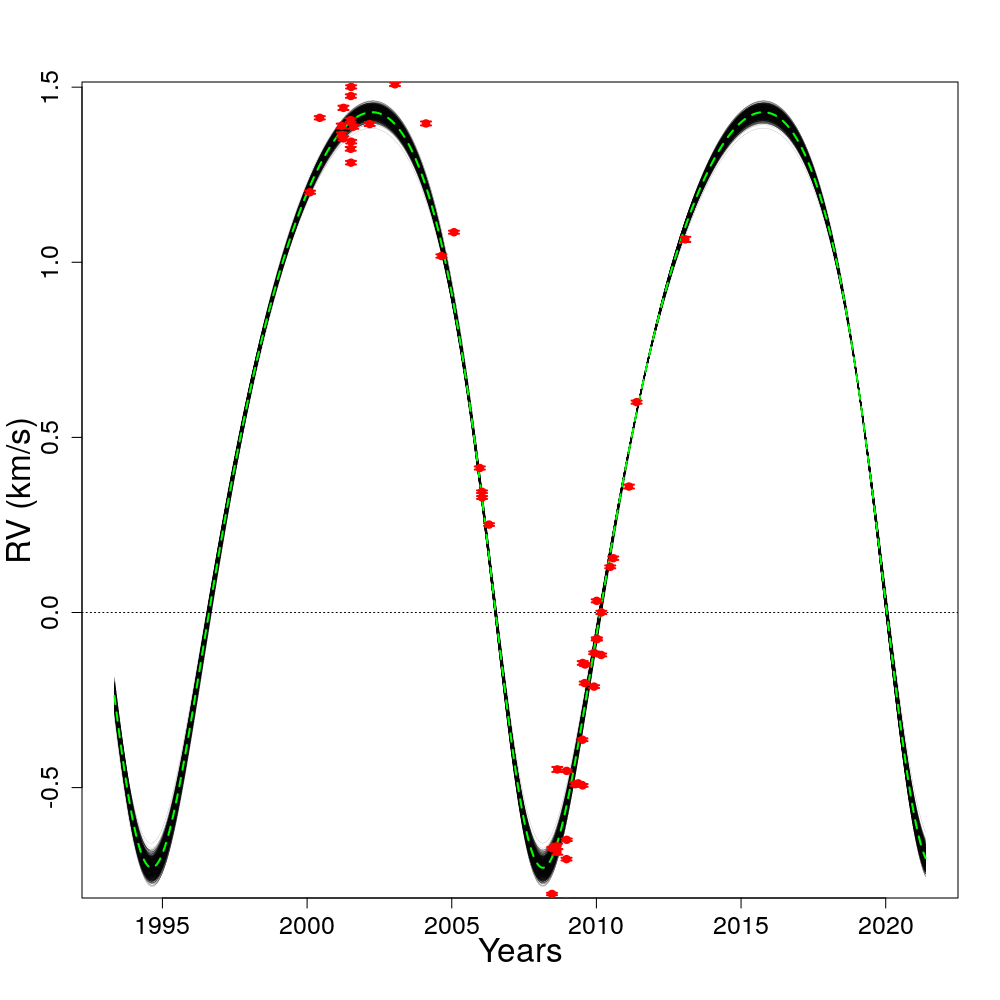}
		\caption{Radial velocity behaviour of Gl~494 predicted by the adjustment of the direct imaging, Hipparcos and Gaia data. The best fit orbit from TMB (values in Table~\ref{OPTable}) is the green dotted line and sample orbits from the MCMC are in black. The red dots represent the radial velocities observations with their associated error bars. The systemic velocity of 0.59 km/s was derived from a TMB adjustment of the radial velocities and the direct imaging data, performed for the sole purpose of this visualization.}
		\label{GL494_2}
	\end{minipage}
\end{figure*}

The Gl~494 (HIP~63510, Ross 458) system around a M star contains a close binary which was first detected astrometrically by \cite{Heintz94} and then resolved with adaptive-optics imaging \citep{ Beuzit2004, Mann2019, Bowler2020}, as well as possibly a common proper motion planetary mass T-dwarf (Gl~494c) at a projected 1200~au (\citealp{Goldman2010} and \citealp{Scholz2010}). We study the inner pair using 16 relative positions compiled by \cite{Bowler2020} and obtained with adaptive optics imagers PUEO on CFHT, NACO on the VLT, and NIRC2 on the Keck telescope, which cover the binary orbit well. 65 radial velocity measurements from the HIRES spectrograph are also available \citep{TalOr2019}, but they are highly impacted by the intrinsic variability of the magnetically active primary star: we only use them for validation purposes as well as to identify which of the two nodes of the orbit is the ascending one. 

The K band contrast ($\Delta m_{K} = 4.27 \pm 0.02$, \cite{Mann2019}) and the much redder spectrum of the late-M secondary guarantee that the secondary star contributes negligible light in the Hipparcos and Gaia observing bands; we will consequently neglect any light from the secondary in the following. The Gaia EDR3 information is indeed compatible with an unresolved source: while the \texttt{ruwe} value is a very high 4.19, the multi peak rate is low, and the Gaia signal is thus compatible with the reflex motion of only one luminous star. We can therefore safely use the published astrometric parameters as representing the average motion of the primary during the first 34 months of the Gaia mission. The solution published in the original reduction of Hipparcos includes an acceleration (7-parameters solution) and the F2 goodness of fit for the 5-parameters solution in the New Reduction of Hipparcos is 2.32, meaning that the reflex motion was already detected by Hipparcos itself. \cite{Kervella2019} also detected a proper motion anomaly, therefore finding signal in the Hipparcos to Gaia difference. 

We adjusted the orbital and astrometric parameters to the relative astrometric data, to the five Gaia astrometric parameters, and to the Hipparcos residual abscissae extracted from the IAD of the new reduction, using the orbital parameters of \cite{Mann2019} as starting values for TMB. Figure~\ref{GL494} represents the relative astrometric observations together with the best orbit obtained with TMB, as well as 1500 MCMC orbit samples to illustrate the uncertainy. Figure~\ref{GL494_2} shows that this adjusted orbit also matches the (unused) radial velocity measurements well, which provides an independent validation. The solution (Table~\ref{OPTable}) is also fully compatible with, but improves upon, the orbits published by \cite{Bowler2020} and \cite{Mann2019}.

%OLD F2=3.26
The goodness of fit of the TMB best solution is F2~=~3.84, and is dominated by two 3~$\sigma$ outliers amongst the relative astrometry observations; if we remove those two, the goodness of fit improves to F2~=~2.01. Our accounting for the reflex motion greatly improves the match to the Hipparcos residual abscissae, with a revised goodness of fit contribution of F2~=~-0.17.

We note that, unsurprisingly for a system with both an orbital period that is $\sim$3 times the length of the Hipparcos mission and a separation of several hundred milliarcseconds, and in agreement with the proper motion anomalies previously detected \citep{MakarovKaplan05,Frankowski07,Kervella2019}, the proper motion that we derive for the barycentre (Table~\ref{APTable}) differs greatly from the published Hipparcos value ($\mu_{\alpha^*}= -616.3 \pm 1.5$ mas/yr and $\mu_\delta= -13.6 \pm 1.0$ mas/yr) and is in full agreement with the long-term proper motion provided in Tycho-2 \citep[$\mu_{\alpha^*}= -640.1 \pm 1.5$ mas/yr, $\mu_\delta= -25.1 \pm 1.4$ mas/yr, ][]{tycho2}. The revised proper motion is much less compatible than the Hipparcos value with the proper motion of the proposed third component C (Table~3 of \citealp{Scholz2010}), with the $\chi^2$ between the proper motions of AB and of C now corresponding to a p-value of $2.62\, 10^{-10}$ instead of $0.01$. We conclude that Gl~494C does not co-move with Gl~494AB and is likely not gravitationally bound to it.

\begin{figure}
\centering
	\includegraphics[scale=0.35]{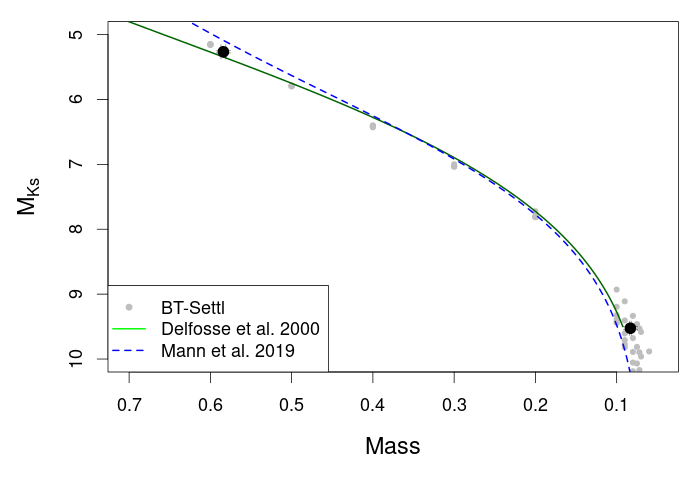}
	\caption{Gl~494 masses compared to the mass-luminosity relations of \cite{Delfosse00}, \cite{Mann2019} and the BT-Settl isochrones \citep{baraffe2015}. \label{fig:Gl494masslum}}
\end{figure}

We also determine, for the first time, the masses of both components of Gl~494 purely from Newtonian physics and without having to adopt a mass of the primary from a Mass-Luminosity relation. Those mass values (also reported in Table~\ref{dynmass}) are $M_1=0.584 \pm 0.003 \Msun$ and $M_2=87 \pm 1 \Mjup$, leading to a total mass for the system of $M_{tot} = 0.667\pm 0.004\Msun$, which is in agreement with a smaller uncertainty with the estimations given by \cite{Mann2019} and \cite{Bowler2020} ($M_{tot}^{\text{\cite{Mann2019}}} = 0.666\pm 0.035 \Msun$ and $M_{tot}^{\text{\cite{Bowler2020}}} = 0.66\pm 0.02 \Msun$). The individual masses we derived are in agreement with the frequently used mass-luminosity relations of \cite{Delfosse00} and \cite{Mann2019} (Fig.~\ref{fig:Gl494masslum}), but the agreement is even better with the BT-Settl isochrones \citep{baraffe2015} in the age range of 150-800 Myr derived by \citealp{Burgasser2010}.

\begin{figure*}
	\begin{minipage}[]{0.48\linewidth}
		\centering
		\includegraphics[scale=0.3]{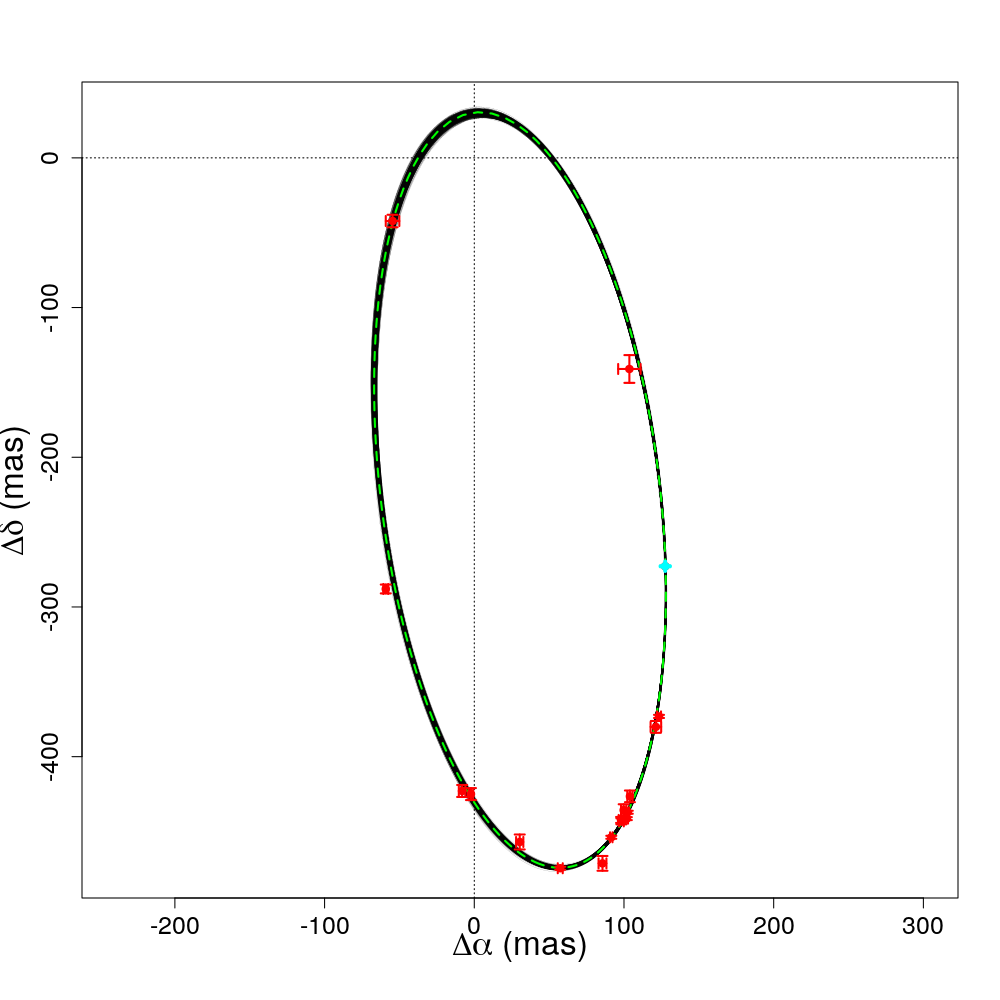}
		\caption{Orbits from the MCMC algorithm in black and of the best solution from TMB in green dotted line for GJ2060 (values in Table~\ref{OPTable}). The direct imaging observations with their associated error bars are in red and the new GRAVITY point with its error bars is in light blue.}
		\label{GJ2060}
	\end{minipage}
	\hfill
	\begin{minipage}[]{0.48\linewidth}
		\centering
		\includegraphics[scale=0.3]{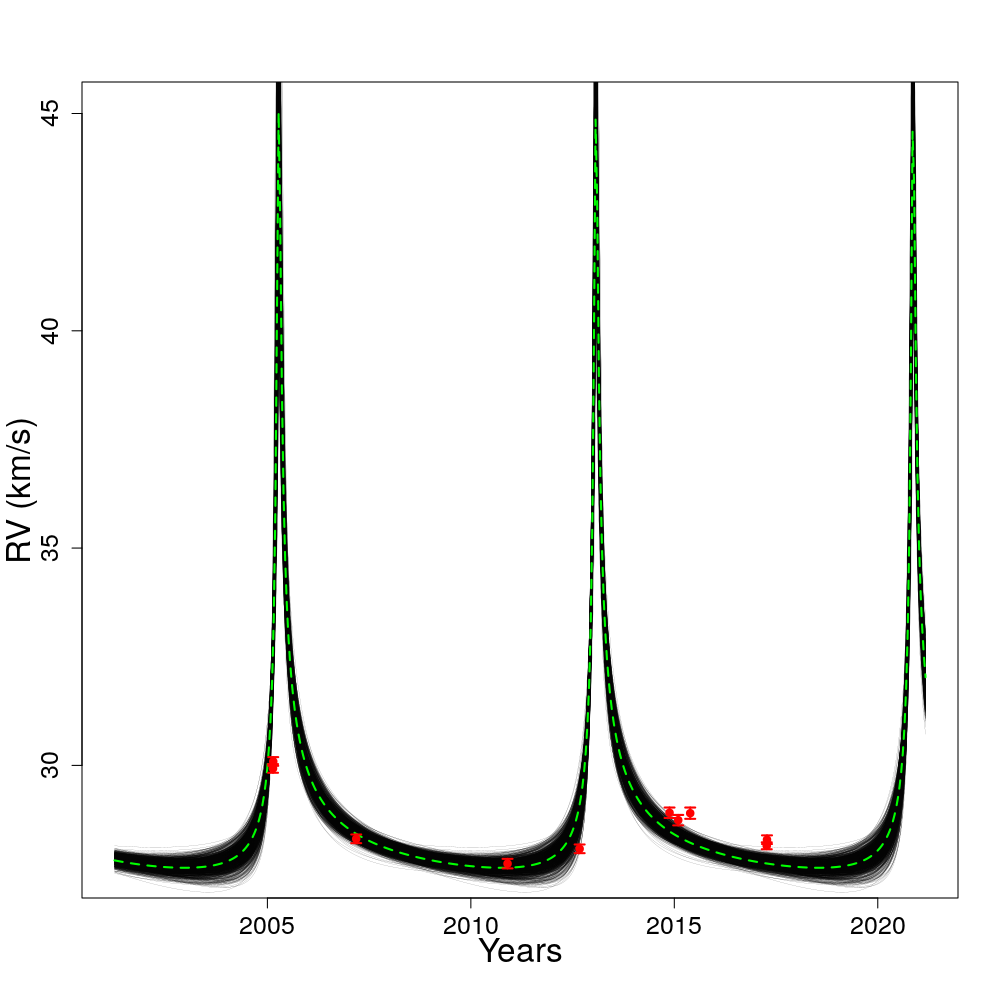}
		\caption{Radial velocity behaviour of GJ2060 predicted by the adjustement of direct imaging and Hipparcos data. The best solution from TMB is the green dotted line (values in Table~\ref{OPTable}) and the MCMC results in black. The radial velocities observations are in red dots with the associated error bars and with $RV_0$=28.8 km/s \citep{Rodet2018}.}
		\label{GJ2060_2}
	\end{minipage}
\end{figure*}

\subsection{GJ~2060: Transit Data and direct imaging}
\label{sec:GJ2060}

The GJ~2060 (HIP~36349) M dwarf system is a member of the AB Doradus moving group whose study is essential for the age prediction of the group that is not well constrained yet (100-150 Myr in the most recent studies: \cite{Barenfeld2013} and \cite{Bell2015}). As such, GJ~2060 adds to the restrained list of young tight binary systems amenable to dynamical measurements. It can vet the evolutionary model predictions known to be impacted by several uncertainties at young ages and in the low-mass regime \citep[see][for a review]{2007prpl.conf..411M}. 

\cite{Rodet2018} previously studied the system using direct imaging and radial velocity observations.
We use the 17 known relative astrometric measurements gathered from multiple imaging instruments (VLT/NaCo, astralux, Gemini/NICI, VLT/SPHERE) \citep{Rodet2018},
 and one new and higher precision one obtained with the VLTI/GRAVITY instrument \citep{2017A&A...602A..94G}. The point is reported in Table~\ref{GRAVITYpoint}.

 %##############################################################
\begin{table}
    %\resizebox{\columnwidth}{!}{
	\caption{New GRAVITY point for the GJ~2060 system.}
        \centering \begin{tabular}{c c c}  
		\hline  Date & $\Delta\alpha$  & $\Delta\delta$ \\  
		(JD-2400000) & (mas) & (mas) \\ \hline \\
	    59623.073 & 127.41 $\pm$ 0.24 & -272.68 $\pm$ 0.33 
		 \\ \\ \hline 
	\end{tabular} 
    %}
	%\hfill
	\label{GRAVITYpoint}
\end{table}
%##############################################################
%\vspace{-0.4cm}

 \cite{Rodet2018} also use 10 radial velocity measurements from FEROS, which we choose to not use since a large jitter is present due to the stellar variability and the measurements are impacted
by the flux of the secondary ($\sim$0.25 flux ratio in the FEROS bandpass). The spectra would ideally be reanalysed as double-lined, but the velocities of the two components are not well separated at any of the FEROS epochs.

Both components of the system contribute significant flux in the Hipparcos (component solution) and Gaia (\texttt{ipd\_frac\_multi\_peak}=76) data, so we have to use the Hipparcos TD rather than the IAD, and it is not analysed as resolved in Gaia EDR3 so we can not use the Gaia astrometric parameters. The F2~=~2.07 in the new reduction of Hipparcos indicates, taking into account the flux of the secondary, a small astrometric signal. 

For this system we adjust the orbit to the relative astrometric observations and to the TD from the new Hipparcos reduction. The photometric variability \citep{Messina2010} is of the same order of magnitude as the Hipparcos photometric data noise, so we did not consider the variability of the primary star. We use the orbital solution of \cite{Rodet2018} as starting values for the TMB gradient descent. A full exploration of the parameter space was also tested leading to the same solution. The solution is represented on the direct imaging data in Fig.\ref{GJ2060}. Figure~\ref{GJ2060_2} shows that this solution is qualitatively consistent with the radial velocity data, which provides an independent validation, and our orbital parameters (Table~\ref{OPTable}) are compatible with \cite{Rodet2018} with the same strong correlation between $\omega$ ad $\Omega$ (Fig.~\ref{fig:GJ2060MCMC}). The goodness of fit of TMB's best solution is F2~=~2.99 and the contribution to F2 of the TD is significantly better for our orbital solution (F2~=~0.76) than the published one (F2~=~2.07). 

\cite{Rodet2018} reported the total system mass from the relative astrometry. Here we directly determine individual dynamical masses of both companions for the first time. The masses derived from our adjusted orbital parameters (Table~\ref{OPTable}) are $M_1=0.61 \pm 0.06 \Msun$, $M_2=0.44 ^{+ 0.06} _{- 0.05}\Msun$ (Table~\ref{dynmass}). The fractional mass deducted is $ \frac{M_2}{M_{tot}}= 0.42 \pm0.04$ and it is consistent with the estimation of \citealp{Rodet2018} ($\frac{m_2}{m_{tot}}= 0.46 \pm 0.10$) from the SB2 assumption using the method proposed by \cite{Montet2015}.
Comparison of the masses and luminosities with stellar evolution models (Appendix~\ref{modelcomparison}) points to an age above 100 Myr, which is consistent with the most recent age estimates for the AB Doradus moving group relying on kinematics and chemistry \citep{Barenfeld2013}, placement of group members on isochrones \citep{Bell2015}, or cosmochronology \citep{2018ApJ...861L..13G}.

\begin{figure*}
	\begin{minipage}[t]{0.48\linewidth}
		\centering
		\includegraphics[scale=0.30]{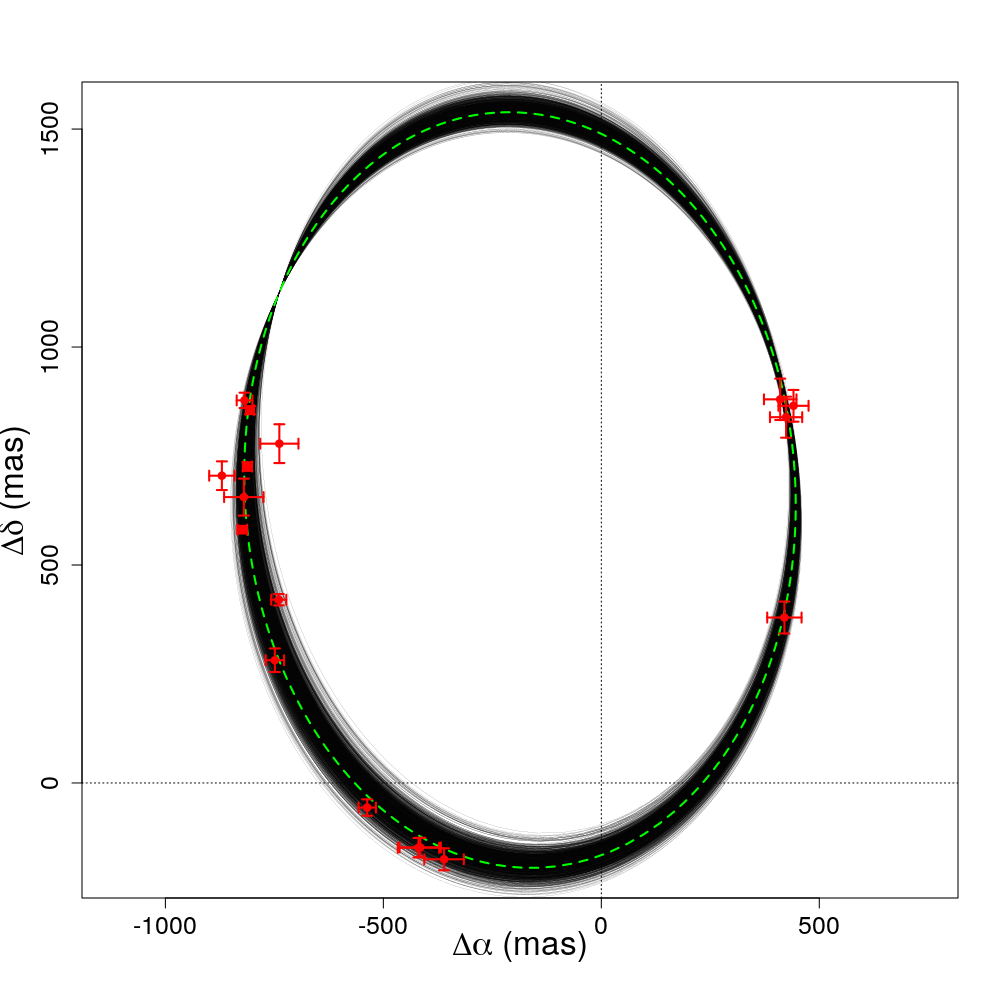}
		\caption{Orbits of HIP~88745 adjusted to the Hipparcos and Gaia data. The best fit TMB solution is displayed as the green dotted line (values in Table~\ref{OPTable}, solution A), and a sampling of the MCMC solutions in black. The direct imaging observations (in red) were not used in the adjustment and provide an independent validation of the orbit.}
		\label{HIP88745}
	\end{minipage}
	\hfill
	\begin{minipage}[t]{0.48\linewidth}
		\centering
		\includegraphics[scale=0.3]{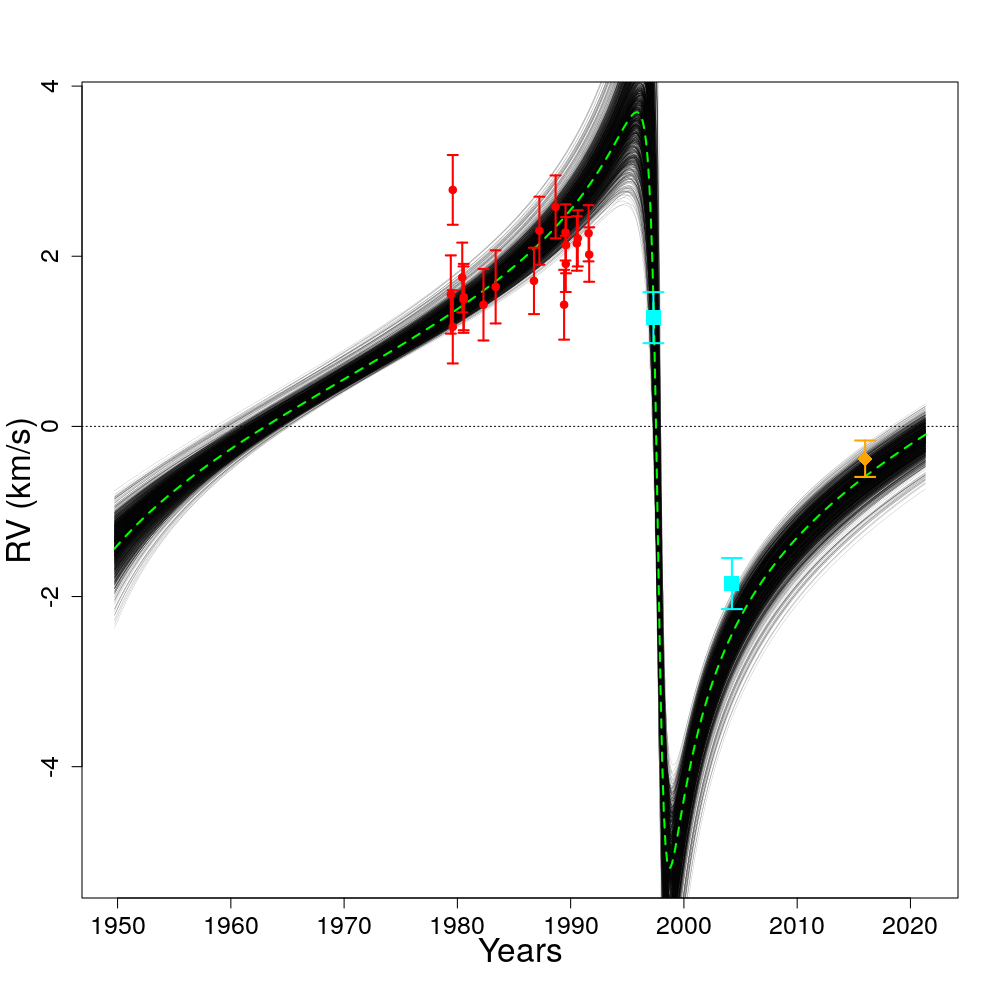}
		\caption{Radial velocity behaviour of HIP~88745 predicted by the adjustment of the Hipparcos and Gaia data. The green dotted line represents the TMB best fit solution (values in Table~\ref{OPTable}, solution A) and the black lines the MCMC solutions. The radial velocities observations from SB9 are in red, the Gaia DR3 radial velocity in orange and the ELODIE archive in blue. The systemic velocity of $RV_0$= 0.40 km/s was determined from an adjustment of the SB9 radial velocities together with TD and Gaia observations which was performed for the sole purpose of this visualisation.}
		\label{HIP88745_2}
	\end{minipage}
\end{figure*}

\subsection{HIP~88745: Hipparcos transit Data and resolved Gaia observation}

HIP~88745 is known to be a binary system with a main-sequence F star \citep{Hutter2019} and a circum-binary polarized debris disk \citep{Kennedy2012}. Direct imaging studies of the system include \cite{Heintz1972}, \cite{Abt2006}, \cite{Kennedy2012}, \cite{Malkov2012} and \cite{Jao2016}. \cite{Soderhjelm1999} studied the system using the TD of the original Hipparcos reduction, and it appears in his Table~4 of stars for which only the total mass of the system could be derived.

We chose to analyse this system because it both has a component solution in Hipparcos and was analysed as resolved in Gaia EDR3. We therefore adjusted to the Hipparcos TD as well as to the Gaia EDR3 astrometric parameters for the primary (5AP$_A$) and secondary (2AP$_B$) components.
The Hipparcos new reduction considered the secondary flux and an astrometric acceleration, but still contains a strong remaining signal (F2~=~9.77). The \texttt{ruwe} value of the 5AP$_A$ Gaia EDR3 solution is 1.39.

Direct imaging and radial velocity data are also available for this star: 17 relative positions from the Fourth Catalog of Interferometric Measurements of Binary Stars \citep{INT4} and 19 radial velocities of the primary from SB9 \cite{SB9} covering a small fraction of the orbital period. To complete those radial velocities, we also consider the Gaia DR3 radial velocity \cite{Katz22} with the error inflation described in \cite{Babusiaux22} and two observations available in the ELODIE archive\footnote{http://atlas.obs-hp.fr/elodie/} \citep{ELODIE}, for which we consider a quite arbitrary 0.3~km/s offset uncertainty, in our validation plots.

For this star we performed two adjustments, one (A) to the absolute astrometry data from Hipparcos and Gaia, and one (AR) that additionally uses the available relative astrometry. The (A) adjustment tests what can be done with a pure absolute astrometry fit and provides parameters that are independent of the direct imaging and radial velocity, and the (AR) adjustment provides better constrained parameters. For both adjustments we used the \cite{Jao2016} orbital parameters as starting values. Tables~\ref{OPTable} and \ref{APTable} present our astrometric and orbital parameters for both adjustments. Figures~\ref{fig:HIP88745MCMC_A} and \ref{fig:HIP88745MCMC_AR} show that, as expected, the correlations are reduced by the introduction of the direct imaging data in the fit. Figures~\ref{HIP88745} and \ref{HIP88745_2} show that the orbital solution from the pure absolute astrometric adjustment matches the direct imaging and radial velocity data well, validating this solution. The fitted 3.81$\pm$0.02~mag contrast between the two components in the Hipparcos band is qualitatively consistent with the Gaia magnitude difference of $\Delta m_G=3.406\pm0.005$~mag, given the bluer Hipparcos passband.
Our adjusted parameters are compatible with \cite{Jao2016} within 3 $\sigma$. 

The global F2 of the (A) and (AR) adjustments are strongly dominated by the Hipparcos TD, for which two 5 $\sigma$ outliers are removed. The global F2 are F2~=~6.97 and 7.1 for the (A) and (AR) adjustments respectively. The new F2 for Hipparcos are F2~=~9.09 and 9.26 for the (A) and (AR) adjustments respectively, which remain equivalent to the published F2 that took into account an astrometric acceleration.
We do not fully understand the reason of this high score, but a third component in the system is one possibility. One was previously listed in the Washington Double Star Catalog \cite{WDS}, before being classified as a non-detection by \cite{Hutter2019}. We note that both components are listed in Gaia EDR3 with non negligible multi peak fraction (31\% and 21\% for the primary and the secondary). For the primary, those might be from transits where the secondary is not separately detected, since the primary has 4 times as many observations used than the secondary.

\begin{figure}
\centering
	\includegraphics[scale=0.5]{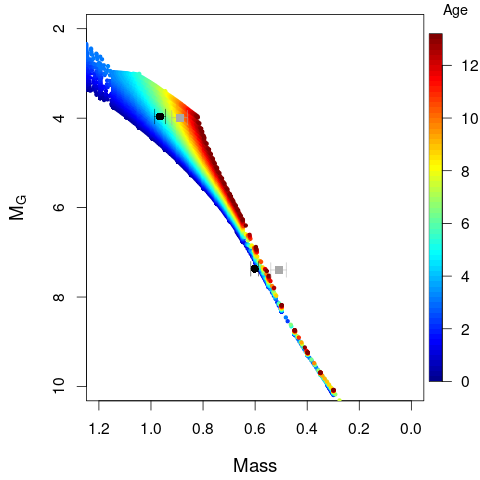}
	\caption{PARSEC isochrones with [M/H]=-0.6~dex colour-coded by age (in Gyr). The dark circles correspond to the masses derived in this work (AR solution) while the grey squares are from \cite{Jao2016}.\label{fig:HIP88745isocs}}
\end{figure}

We directly determine individual dynamical masses (Table~\ref{dynmass}) of both companions using the TD and Gaia astrometric parameters of the primary and secondary. The masses derived from the adjusted orbital parameters (Table~\ref{OPTable}) using absolute astrometry only are $M_1=1.01 \pm 0.04 \Msun$, $M_2=0.68 \pm 0.04 \Msun$ and using absolute astrometry and relative astrometry $M_1=0.96 \pm0.02\Msun$ and $M_2= 0.60^{+ 0.02 } _{- 0.01 } \Msun$. These values are at around $3\sigma$ from the estimations made by \cite{Jao2016} from the SB1 mass function and the relative and photocentre orbit,  $M_1^{\text{\cite{Jao2016}}}=0.89\pm0.03 \Msun$ and $M_2^{\text{\cite{Jao2016}}}=0.51\pm0.03 \Msun$. 
Figure~\ref{fig:HIP88745isocs} shows that our new masses are more in agreement with the PARSEC isochrones \citep{bressan2012} for a metallicity of -0.6~dex \cite[see references in ][]{Jao2016} than the \cite{Jao2016} masses. The primary star mass would lead to an age of the system of 5$\pm$1.3~Gyr for a -0.6~dex metallicity.

\begin{table*}[htp]
		\caption{Orbital parameters adjusted for Gl~494, GJ~2060 and HIP~88745. The parameters are estimated with TMB and their errors estimated thanks to a MCMC. HIP~88745$^{(A)}$: using only absolute astrometry. HIP~88745$^{(AR)}$: using absolute and relative astrometry.} 
		\label{OPTable}
		\centering 
		%\small
		\resizebox{2\columnwidth}{!}{
		\begin{tabular}{ c| c c c c c c c c c c c c}  
		\hline OP & P & Tp & $a_1$ & e  & $\omega_2$ & i & $\Omega$ & $a_{21}$ & $\Delta m_{HIP}$  \\ 
		  & Years & (days, J2000) & (a.u) & & (deg) & (deg) & (deg) & (a.u) &  \\ \hline \hline \\
		Gl~494 & 13.52 $\pm0.02 $ & 7721 $\pm8 $ & 0.62 $^{+ 0.009 } _{- 0.008 }$ & 0.243 $\pm0.001 $ & 336.3 $\pm0.4$ & 130.0 $\pm0.1$ & 236.7 $\pm 0.1$ & 4.959 $\pm0.009$ & $ \displaystyle \times$ \\ \\ \hline \\ 
		
		GJ~2060 & 7.794 $\pm 0.008$ & 1926 $^{+ 7 } _{- 6 }$ & 1.7 $ \pm 0.2$ & 0.882 $^{+ 0.004 } _{- 0.005 }$ & 169 $\pm3$ & 40 $\pm1$ & 180 $\pm3$ & 4.0 $\pm0.1$ & 1.93 $\pm0.05$ \\ \\ \hline \\
		
		HIP~88745$^{(A)}$ & 60 $^{+ 3 } _{- 2 }$ & -803 $^{+ 30 } _{- 33 }$ & 7.3 $^{+ 0.6 } _{- 0.5 }$ & 0.82 $\pm 0.02$ & 285 $\pm 2$ & 44 $\pm 2$ & 234 $\pm3$ & 18.2 $^{+ 0.8 } _{- 0.6 }$ & 3.81 $\pm 0.01$ 
		
		 \\ \\ \hline 
		  \\
		 HIP~88745$^{(AR)}$ & 56.5 $\pm 0.4$ & -761 $^{+ 20 } _{- 16 }$ & 6.56 $\pm 0.06$ & 0.783 $\pm 0.003$ & 288.8 $\pm 0.5$ & 40.3 $\pm 0.5$ & 229.8 $^{+ 0.8 } _{- 0.7 }$ & 17.1 $\pm 0.1$ & 3.81 $^{+ 0.02} _{- 0.01 }$
		 
		 \\ \\ \hline 
		\end{tabular} 
		}
		\hfill	
	\end{table*}

\begin{table*}[htp] 
	\caption{Astrometric parameters at Hipparcos reference adjusted for Gl~494, GJ~2060 and HIP~88745. The parallax takes into account the zero-point of Hipparcos parallaxes. The parameters are estimated with TMB and the errors are estimated thanks to a MCMC. HIP~88745$^{(A)}$: using only absolute astrometry. HIP~88745$^{(AR)}$: using absolute and relative astrometry.}
	\label{APTable}
    %\resizebox{2\columnwidth}{!}{
        \centering 
        %\small
        \begin{tabular}{c|c c c c c }  
		\hline  Astrometric & $\alpha$ & $\delta$ & $\varpi$ & $\mu_{\alpha^*} $ & $\mu_\delta$  \\ 
		parameters & (deg $\pm$ mas) & (deg $\pm$ mas) & (mas) & (mas/yr) & (mas/yr) \\ \hline \hline \\
		Gl~494 & 195.1956345 $\pm 1 $ & 12.3757745 $\pm0.9 $ & 86.6 $\pm0.1$ & -638.63 $^{+ 0.05 } _{- 0.06 }$ & -24.80 $\pm 0.04$ \\ \\ \hline \\
		GJ~2060 & 112.2144281 $\pm4$ & -30.2465342 $\pm11$ & 64 $^{+ 1 } _{- 2 }$ & -126 $\pm1$ & -182 $\pm5$  \\  \\ \hline \\
		HIP~88745$^{(A)}$ & 271.7567522 $^{+ 6 } _{-5}$ & 30.5619679 $^{+ 13 } _{- 12 }$ & 63.52 $^{+ 0.08 } _{-0.09}$ & -92.5 $\pm0.6$ & 73.3 $\pm 0.1$ \\  \\ \hline \\
		HIP~88745$^{(AR)}$ & 271.7567501 $\pm1$ & 30.5619630 $\pm2$ & 63.54 $\pm 0.08$ & -91.7 $\pm0.1$ & 73.20 $\pm 0.05$
		\\  \\ \hline 
	    \end{tabular} 
    %}
	\hfill
\end{table*}

\begin{table*}[htp]
    \caption{Summary of the dynamical masses adjusted in this paper for the 3 systems Gl~494, GJ~2060 and HIP~88745 (both absolute astrometry only and absolute and relative astrometry adjustments).} 
		\label{dynmass}
		\centering 
		%\small
		%\resizebox{2\columnwidth}{!}{
		\begin{tabular}{ c| c c c c}  
		\hline  & Gl~494 & GJ~2060 & HIP~88745$^{(A)}$ & HIP~88745$^{(AR)}$ \\ \hline \hline \\
		Primary mass  & $M_1=0.584 \pm 0.003 \Msun$ & $M_1=0.60 ^{+ 0.06} _{- 0.05} \Msun$ & $M_1=1.01 \pm 0.04 \Msun$ &  $M_1=0.96 \pm0.02\Msun$ \\ \\
		Secondary mass &  $M_2=87 \pm 1 \Mjup$ &$M_2=0.45 ^{+ 0.06} _{- 0.05}\Msun$ & $M_2=0.68 \pm 0.04 \Msun$ & $M_2= 0.60^{+ 0.02 } _{- 0.01 } \Msun$
		 \\ \\ \hline 
		\end{tabular} 
		%}
\end{table*} 

\section{Discussion and conclusion}

\label{section:conclusion}
We presented our new BINARYS tool which rigorously combines Hipparcos and Gaia observations of binary stars with relative astrometry and/or radial velocity observations. For systems where the secondary contributes significant light, BINARYS uses the raw Hipparcos transit data.

For illustration and validation, we presented 3 systems studied with BINARYS. The adjustment of direct imaging, Hipparcos IAD, and Gaia EDR3 constraints the primary and secondary masses in the Gl~494 system to $M_1=0.584 \pm 0.003 \Msun$ and $M_2=87 \pm 1 \Mjup$. That adjustment also indicates that Gl~494C is unlikely to co-move with Gl~494AB. 

The adjustment of direct imaging and Hipparcos TD on the AB Doradus GJ~2060AB system determines the masses of its primary and secondary $M_1=0.60 ^{+ 0.06} _{- 0.05} \Msun$ and $M_2=0.45 ^{+ 0.06} _{- 0.05}\Msun$, which in turn constrains the age of the system to older than 100~Myr, in good agreement with the most recent estimate of the moving group age.

Finally, the adjustment of Hipparcos TD and resolved Gaia observations of HIP~88745 gave masses for the primary and secondary of $M_1=0.96 \pm 0.02\Msun$ and $M_2= 0.60^{+ 0.02 } _{- 0.01 } \Msun$, with strong residuals in the Hipparcos TD. Those may reflect a potential remaining signal in the TD, which might become usable later with further information and which could be from a third component. 
	
%perspectives
In the future, we plan to extend BINARYS to accommodate very nearby stars which have significant perspective acceleration during the Gaia and Hipparcos missions, and stars in clusters, where light from a star outside the system can contaminate Hipparcos observations. BINARYS is also being extended for the study of triple system \citep{Lagrange2020} and to take into account the new non-single solutions (NSS) that are provided by Gaia DR3 \citep{Arenou22}.

%perspectives Gaia
Additionally, this tool prepares for Gaia DR4, which will provide epoch observations. At that point, we will be able to combine Hipparcos and the Gaia equivalent of the TD. The tool will by then run mostly without Hipparcos constrains due to the huge sample size difference. Although non-single solutions will be provided by the Gaia-DPAC consortium, the combination of Gaia with external data will have to be done on the epoch data for an optimized solution, but also to derive solutions for systems with a too faint Gaia signal to have a full NSS solution and to handle specific cases such as multiple systems. The fine and accurate handling of the Gaia epoch data will be crucial for the study of the exoplanets expected to be discovered by Gaia, up to $\sim$70,000 for a 10 years mission \citep{Perryman2014}.

\begin{acknowledgements}
We thank the referee for his detailed comments that helped to improve the clarity of the manuscript.
This work is supported by the French National Research Agency in the frame-work of the Investissements d’Avenir program (ANR-15-IDEX-02), in particular through the funding of the "Origin of Life" project of the Univ. Grenoble-Alpes.
This work has made use of data from the European Space Agency (ESA) mission {\it Gaia} (\url{https://www.cosmos.esa.int/gaia}), processed by the {\it Gaia} Data Processing and Analysis Consortium (DPAC, \url{https://www.cosmos.esa.int/web/gaia/dpac/consortium}). Funding for the DPAC has been provided by national institutions, in particular the institutions participating in the {\it Gaia} Multilateral Agreement.
\end{acknowledgements}

\bibliographystyle{aa} % style aa.bst
\bibliography{biblio}

%%%%%%%%%%%%%%%%%%%%%%%%%%%%%%%%%%%%%%%%%%%%%%%%

\appendix

\section{Model comparison for GJ2060\label{modelcomparison}}

Let us now use the dynamical masses obtained in Section~\ref{sec:GJ2060} for the GJ2060 system to derive the age of the stars, and thus increase the constraints on the age of its young moving group ABDor. We retrieve from \cite{Rodet2018} the bolometric luminosities $L$ of each star. They were derived using a distance $d = 15.69 \pm 0.45$~pc, which is compatible with the parallax that we obtain in this work (Table~\ref{APTable}). Since the binary is young, we use pre-main sequence (PMS) evolutionary models from the literature to relate mass, luminosity and age.

Several evolutionary models for PMS stars rely on slightly different physics (e.g., atmospheric models, convection efficiency). We used models from \citet[hereafter BHAC15]{baraffe2015}, \citet[hereafter DM97]{Antona1997}, the PARSEC model \citep{bressan2012}, the PISA model \citep{tognelli2011,tognelli2012}, the Darmouth model \citep{dotter2008,feiden2014} and the one from \citet[hereafter Siess00]{siess2000}. When the model requires stellar parameters (hydrogen, helium or metal composition), we used the ones closest to the solar abundances \citep[as given in][]{asplund2009}. Such hypotheses are consistent with the solar-like metallicity derived for members of the ABDor moving group \citep{mccarthy2014}.

\begin{figure}
	\centering
	\includegraphics[scale=0.5]{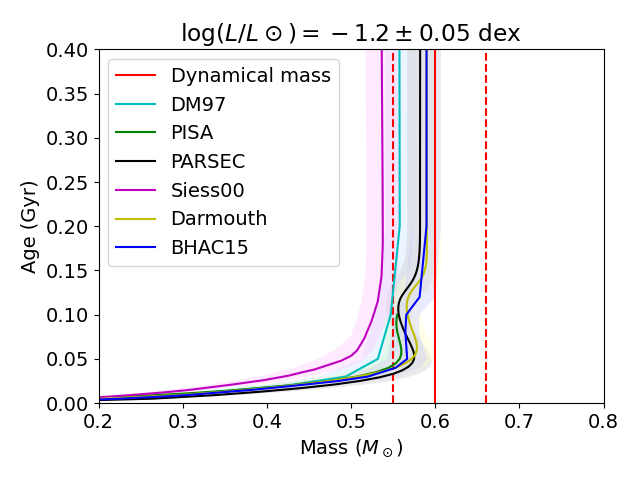}
	\includegraphics[scale=0.41]{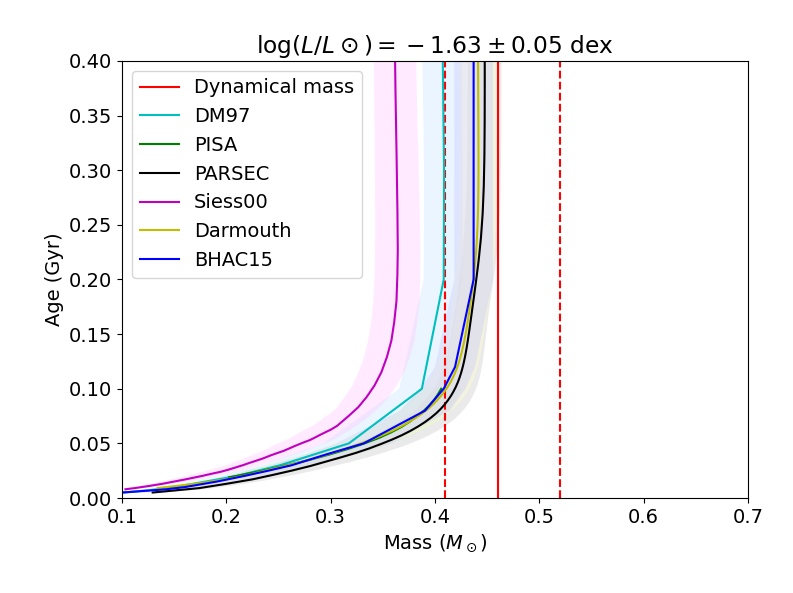}
	\caption{Primary (top) and secondary (bottom) mass of the GJ2060 system compared with mass-age relations coming from six evolutionary models. The red vertical lines correspond to the mass estimates derived in this work. The shades correspond to the uncertainty in the luminosities. The masses and luminosities suggest that the system is older than 100 Myr.}
	\label{fig:gj2060models}
\end{figure}

We plot the masses as a function of the system age for the given luminosity in Figure~\ref{fig:gj2060models}. The shade shows the uncertainties associated with the error on the luminosity. Pre-main sequence low-mass stars are more luminous than their main-sequence counterparts, so that a given luminosity can correspond to both a young low-mass star or an older more massive star. The plot diverges at the main-sequence mass corresponding to the observed luminosity. Indeed, the luminosity evolves on much larger timescales when the star reaches the zero-age main-sequence (at around 100 Myr old), so that all ages greater than 100 Myr are roughly compatible with the main-sequence mass.

The discrepancy with the models is reduced compared to the study of \cite{Rodet2018}, due to the slightly lower masses that we derived in this work. Our values are now compatible with most of the models assuming the system is at least 100 Myr old. This age agrees with recent independent estimates of the ABDor moving group, arguing for its similarity with the $\sim$120-Myr Pleiades. However, our results are not compatible with the predictions from the Siess00 model, and only marginally compatible with the predictions from DM97.

\onecolumn 
\newpage

\section{Corner plots}
\begin{figure}[htp]
	\centering
	\includegraphics[scale=0.7]{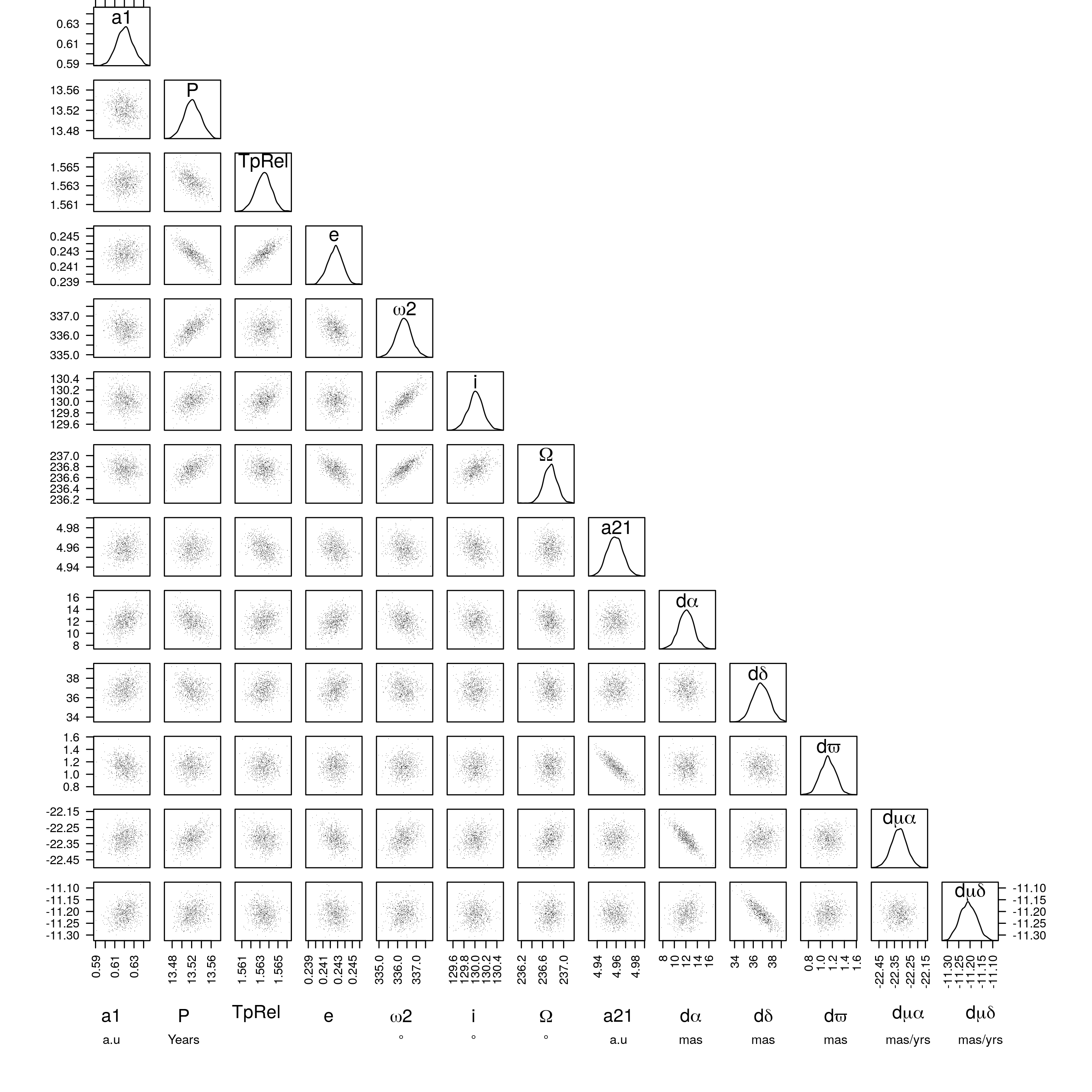}
	\caption{Corner plot of all the MCMC iterations for Gl~494 that shows the correlations between parameters and their density of probability diagonally. The delta values given for the 5 astrometric parameters are given with respect to the solution published in the IAD.\label{fig:GL494MCMC}}
\end{figure}

\begin{figure}[htp]
	\centering
	\includegraphics[scale=0.7]{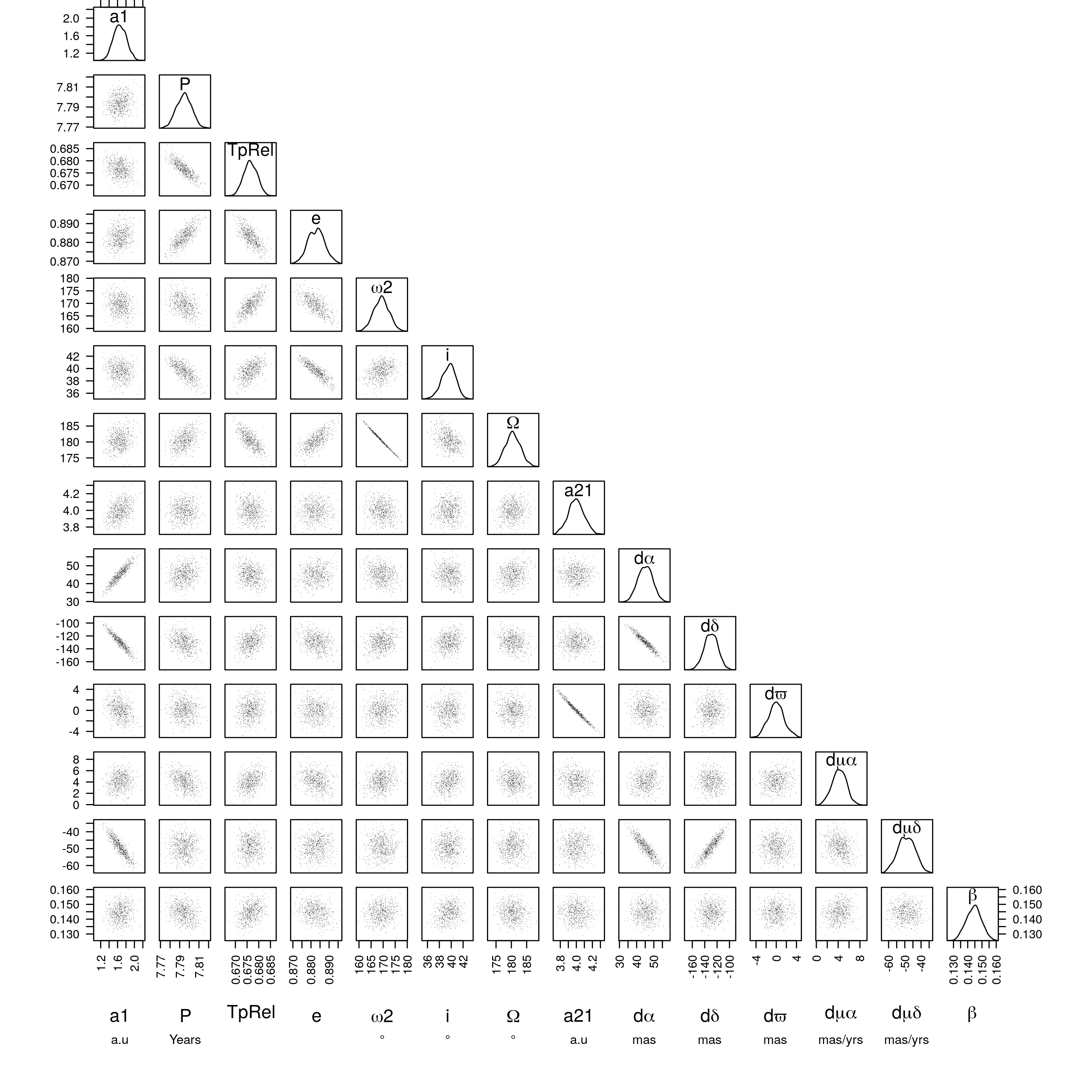}
	\caption{Same as Fig.~\ref{fig:GL494MCMC} for GJ~2060.\label{fig:GJ2060MCMC}}
\end{figure}

\begin{figure}[htp]
	\centering
	\includegraphics[scale=0.7]{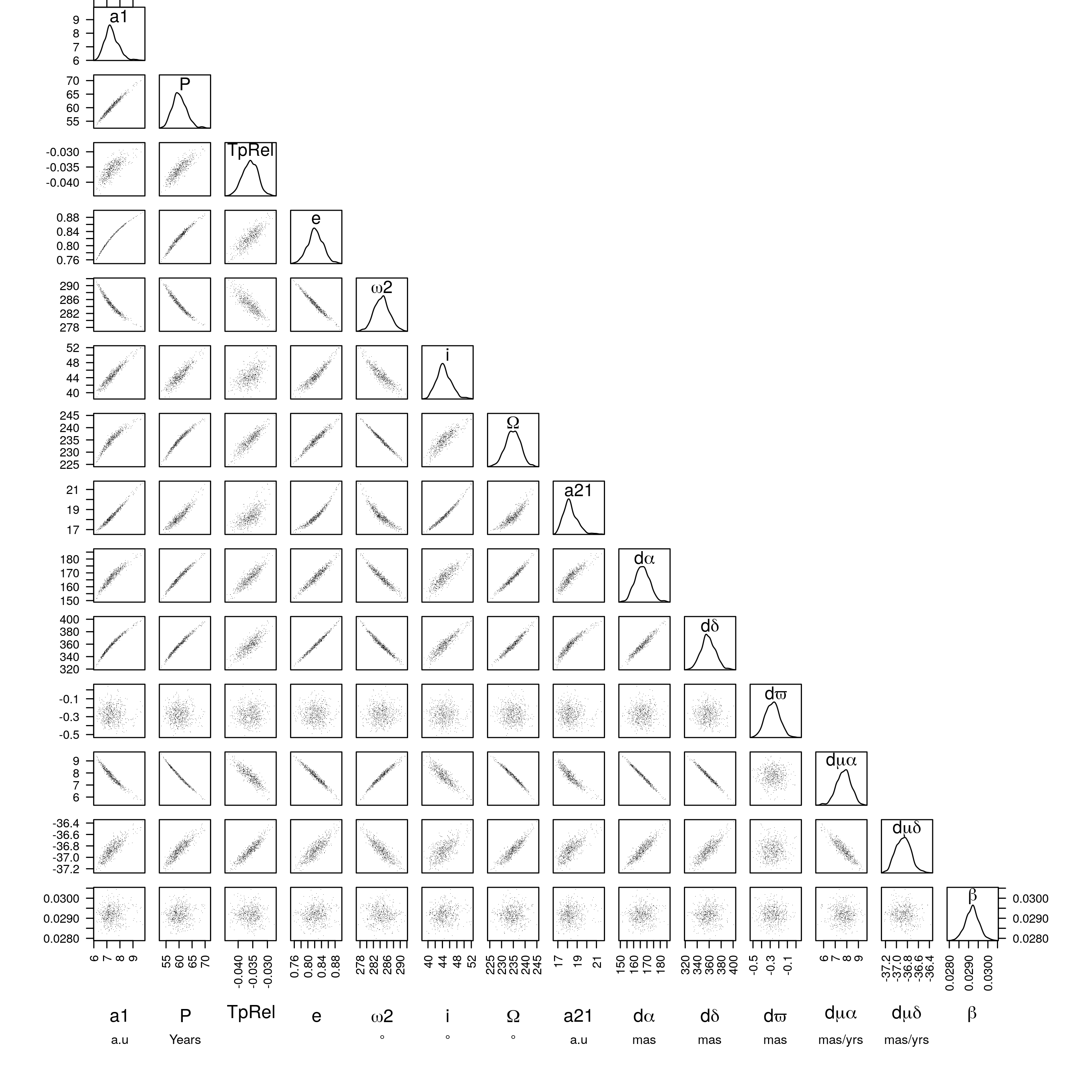}
	\caption{Same as Fig.~\ref{fig:GL494MCMC} for HIP~88745 adjustment (A) using  only the absolute astrometry from Hipparcos and Gaia.
	\label{fig:HIP88745MCMC_A}}
\end{figure}

\begin{figure}[htp]
	\centering
	\includegraphics[scale=0.7]{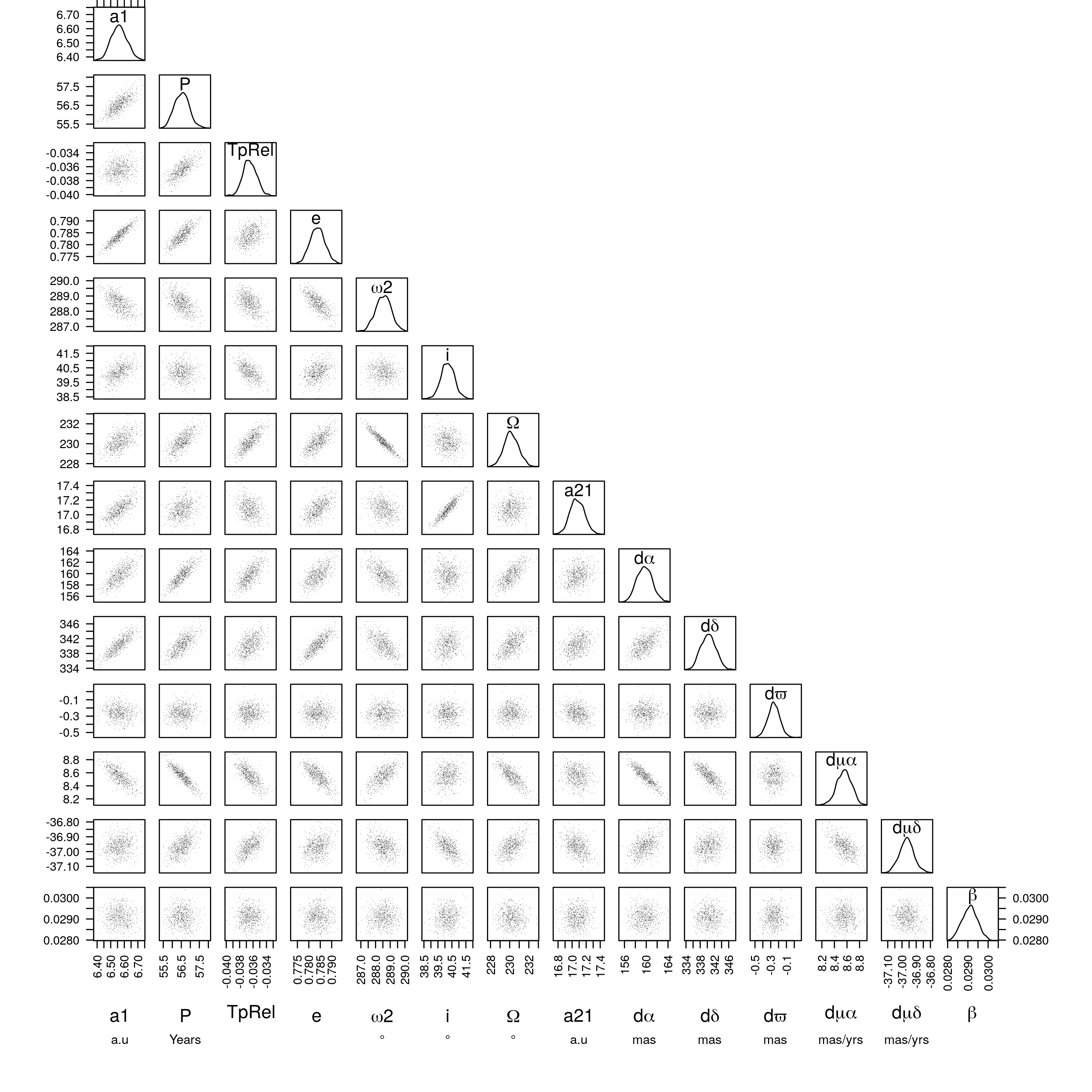}
	\caption{Same as Fig.~\ref{fig:GL494MCMC} for HIP~88745 adjustment (AR) using  both absolute and relative astrometric data.
	\label{fig:HIP88745MCMC_AR}}
\end{figure}

\end{document}